# FERROELECTRIC 180 DEGREE WALLS ARE MECHANICALLY SOFTER THAN THE DOMAINS THEY SEPARATE

Christina Stefani[1], Louis Ponet[2,3], Konstantin Shapovalov[4], Peng Chen[2], Eric Langenberg[5], Darrell G. Schlom[5,6] Sergey Artyurhin[2], Massimiliano Stengel[4,7], Neus Domingo[1], Gustau Catalan[1,7]

[1] *ICN2-Institut Català de Nanociència I Nanotecnologia (CERCA-BIST-CSIC), Campus Universitat Autònoma de Barcelona, Bellaterra, Catalonia*
[2] *Italian Institute of Technology, 16163 Genoa GE, Italy*
[3] *Scuola Normale Superiore di Pisa, 56126 Pisa PI Italy*
[4] *ICMAB-Institut de Ciència de Materials de Barcelona, Bellaterra, Catalonia*
[5] *Department of Materials Science and Engineering, Cornell University, Ithaca, New York 14853, USA*
[6] *Kavli Institute at Cornell for Nanoscale Science, Ithaca, New York 14853, USA*
[7] *ICREA-Catalan Institution for Research and Advanced Studies, Passeig Lluïs Companys, Barcelona, Catalonia*

**Abstract.**
Domain walls are functionally different from the domains they separate, but little is known about their mechanical properties. Using scanning probe microscopy, we have measured the mechanical response of ferroelectric 180º domain walls and observed that, despite separating domains that are mechanically identical (non-ferroelastic), the walls are mechanically distinct -softer- compared to the domains. This effect has been observed in different ferroelectric materials (LiNbO$_3$, BaTiO$_3$, PbTiO$_3$) and with different morphologies (from single crystals to thin films) so it appears to be universal. We propose a theoretical framework that explains the domain wall softening and justifies that the effect should be common to all ferroelectrics.

## I. Introduction.

An important part of the appeal of domain walls resides in the functional contrast between their properties and those of the domains they separate. Multiferroic BiFeO$_3$ displays conductivity and magnetoresistance at its domain walls[1–4] despite the bulk being an insulator, and the [5]ferroelastic domain walls of semiconductor WO$_{3-x}$ are superconducting[6]. Electrical conductance has also been measured in the ferroelectric domain walls Pb(Zr$_{0.2}$Ti$_{0.8}$)O$_3$[7], LiNbO$_3$[8] and BaTiO$_3$, as well as those of multiferroic YMnO$_3$[9] and Cu$_3$B$_7$O$_{13}$Cl[10]. Their distinct functionality, nanoscopic thickness, and the fact that they can be manipulated, created or moved by an external field is fueling the field of "domain wall nanoelectronics", where domain walls are regarded as mobile two-dimensional electronic elements[11–13].

In contrast to the vigorous research on domain wall functionality, less is known about their mechanical properties. This is partly explained by the difficulty of isolating the mechanical response of individual walls, being atomically thin structures sandwiched between much wider domains that dominate the overall mechanical behavior. All the same, the mechanical response of domain walls matters. For one thing, mechanical stress is one of the mechanisms by which domain walls can be moved: ferroelectric-ferroelastic domain walls respond to stress[14] and/or to electric fields, affecting the overall dielectric, piezoelectric and elastic properties of ferroelectric ceramics as well as their lability[15] and fracture physics [16,17]. Even purely ferroelectric (i.e., non-ferroelastic) domain walls also react to strain gradients introduced by external indentation[18] or by the proximity of another ferroelectric (non-ferroelastic) domain wall[19].

The interplay between domain wall motion/domain reconfiguration and the overall mechanical and electromechanical properties of ferroelectric devices is therefore well documented[20–23]. In contrast, there is barely any knowledge of the internal deformation mechanics of the individual walls themselves –particularly for non-ferroelastic 180º walls separating antiparallel ferroelectric domains. A seminal investigation by Tsuji et al. suggested that the 180º domain walls of ferroelectric PZT ceramics appear softer than the domains when probed by atomic force microscopy [24,25]. However, it is not obvious why such domain walls should display any mechanical contrast, given that the polar axis is the same on both sides of the wall and thus the domains on either side are mechanically identical (unlike in ferroelastic 90º domains, where the spontaneous strain axis is different, so the mechanical properties must necessarily change across the wall). The present investigation therefore seeks to (i) determine whether domain wall softening is a general property of 180º ferroelectric domains, (ii) quantify the magnitude of this wall softening and (iii) propose a theoretical explanation for its physical origin. We find that the effect is general, quantitatively significant, and physically inevitable.

Besides its fundamental interest, this discovery has practical ramifications not only for the mechanics of ferroelectrics but also for their functionality. In particular, heat transport is intimately linked to mechanics because heat is carried by phonons, which are strain waves. The elastic contrast between domains and domain walls is therefore likely to affect the propagation and scattering of phonons –and, consequently, also the propagation of heat[26,27,28]. More specifically, if the lattice is softer at the wall, the phonon speed will be slower, and hence effects like phonon refraction or even total internal reflection may be expected; in this context, 180º domain walls could conceivably act as "phonon waveguides" where heat would travel along the wall with little or no dissipation. Since ferroelectric 180º walls can be created or destroyed by voltage (by writing or erasing domains), this suggests the possibility of using voltage to fabricate periodic and reconfigurable metamaterials with a regular pattern of internal elastic contrast. Put another way: periodically poled ferroelectric crystals, which are already in use for photonic applications[29], may also turn out to be phononic crystals.

### II. Samples and domain structure

We have characterized the mechanical properties of domain walls in ferroelectric single crystals of $LiNbO_3$ and $BaTiO_3$, and thin films of $PbTiO_3$. The spread of materials and sample morphologies was chosen in order to determine the generality of the findings. The measurements were based on Contact Resonance Frequency Microscopy mode (CR-FM).

CR-FM a scanning probe microscopy technique that maps, with nanoscopic resolution, the resonance frequency of an AFM tip in contact with the material; higher resonance frequencies correlate with stiffer contact areas and, conversely, lower resonance frequencies indicate that the material is softer[30]. As we shall show, besides imaging, this technique can be used to extract quantitative information about differences in Young's modulus of the material.

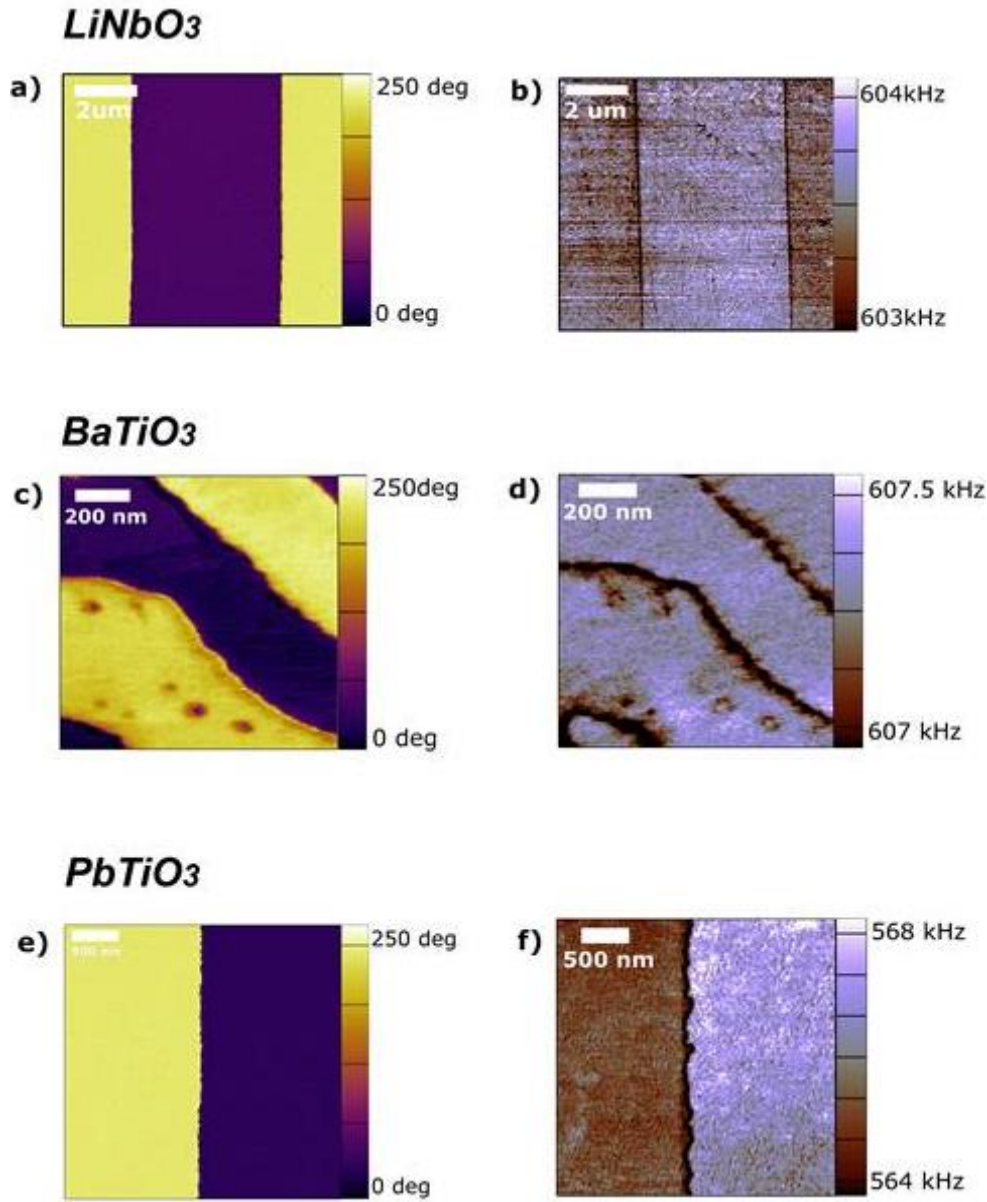

**Figure 1**.(a,b) Periodically poled LiNbO$_3$ single crystal, (c,d) BaTiO$_3$ single crystal spontaneously polarized and (e,f) PbTiO$_3$ thin film. PFM images (a,c,e) showing the opposite polarization of out of plane domains and CR-FM images (b,d,f) demonstrate changes in frequency between domains and domain walls.

Figure 1 shows the Piezoresponse Force Microscopy (PFM)[31,32] images of the ferroelectric domains, showing the 180° degree phase contrast of oppositely polarized domains. Figure 1 also demonstrates the mechanical response of domain walls as detected by Contact Resonance Force Microscopy (CR-FM) mode. In this technique, the tip is in contact with the sample, but it does not excite the sample electrically, as in PFM, but mechanically, through a piezo element placed at the base of the cantilever. This mechanical excitation induces vibration to the sample, and the frequency of the oscillation is modulated until mechanical resonance is reached. The resonance frequency of the system depends on both the geometrical characteristics of the tip and the tip-sample mechanical contact characteristics. Since the tip is the same throughout the whole measurement, and the force between tip and sample is also kept constant, changes of resonant frequency therefore indicate the changes in the stiffness of the sample, and a lower resonance frequency means a softer material. In all the measurements, non-coated doped Si tips were used. For CR-FM images, the tip was grounded during the measurements in a short-circuit configuration, allowing for polarization charge screening.

The first sample tested was a congruent $LiNbO_3$ single crystal, periodically poled with polarization perpendicular to the surface. This material is a uniaxial ferroelectric, so the only domains allowed by symmetry are 180° domains of antiparallel polarization. The antiparallel domain configuration was verified by the phase contrast in the PFM image (Figure 1a). The size of each domain is ~4μm. From CR-FM response of the same area, we observe that the domain walls are markedly darker (i.e., display lower resonance frequency and are therefore softer) than the domains.

We also looked at a $BaTiO_3$ single crystal, which is considered an archetype of perovskite ferroelectrics. The tetragonal structure of $BaTiO_3$ allows for both antiparallel (180°) and perpendicular (90°) domain configurations, the latter being ferroelastic in addition to ferroelectric. In Fig 1b, the PFM image shows the areas with opposite out of plane polarization forming 180° ferroelectric domain walls. In the same area, CR-FM measurement demonstrates again a downward frequency shift between domain and domain walls, indicating that in this material domain walls are also mechanically softer that domains.

Finally, we also investigated a $PbTiO_3$ thin film of 50 nm thickness epitaxially grown by reactive molecular beam epitaxy on a single-crystal $SrTiO_3$ substrate (the growth details can be found elsewhere[26]). Due to the large compressive stress exterted by the substrate (-1.36%), only domains with vertical (out of plane) polarization are allowed in the ferroelectric film[26]. After electrically polarizing two different areas of the film with a dc voltage of 5V, we poled domains of opposite sign, as indicated by the corresponding PFM phase contrast (Fig 1c). Again, the 180° domain walls are observed to be softer than the surrounding domains (Fig 1e). The relative softness of ferroelectric domain walls therefore appears to be a general phenomenon that does not rely on composition or sample morphology.

The observed softening of 180-degree domain walls is qualitatively similar for all the samples, irrespective of whether the domains were artificially written, such as in PbTiO3 or LiNbO3, or spontaneous, as in the BaTiO3 crystal.
It has been proposed that some ferroelectric domain walls can be Neel-type and thus have an in-plane component of the polarization have [33,34]. We have no experimental evidence for this being the case in the BaTiO3 walls, but the possibility of a mechanical contribution coming from in-plane polarization at the walls is excluded because in-plane polarization leads to stiffer, not softer, response (see higher resonance frequency of the a-domain).

The possibility of softening due to local switching effects could also be ruled out. Although it is expected that the coercive field of the ferroelectric should be smaller near the ferroelectric wall[35], repeated scans over the same area show no evidence of switching of the polarization –there is no detectable shift in the position of the domain walls even after 10 scans with the maximum mechanical load of 20 micro-Newtons. There can be, however, a temporary deflection of the wall towards the tip that might be regarded as a "temporary switching", and this is the basis for the theoretical model described in Figure 4.

It is worth mentioning that there is also a small mechanical contrast (small difference in CR-FM) between up and down domains. This contrast is attributed to the coupling of tip-induced flexoelectricity and domain ferroelectricity, which induces an asymmetry in the mechanical response of domains of opposite polarity[36–38]. The domain walls, however, are softer than either up or down-polarized domains, so their softening cannot be explained by this polarity-dependent mechanism.

### III. Experimental quantification of domain wall elasticity

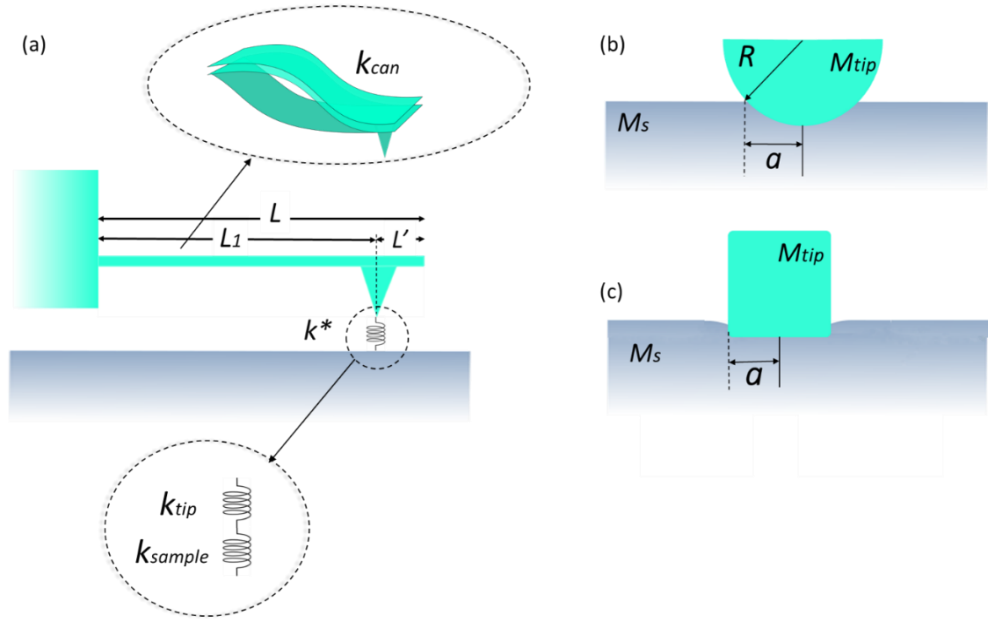

Figure 2. (a) Schematic presentation of AFM cantilever and tip-sample contact being simulated by a spring with constant k*. (b) Diagram of tip- sample contact based on Hertzian mechanical contact model and (c) flat punch contact between the tip and the sample

In order to quantify the softening, we need to translate the shifts in resonance frequency into changes of stiffness. The system can be described as a system of two springs in series (Fig.2a): the cantilever itself, with its flexural elastic constant, and the tip-surface contact, which can be described by the Hertzian contact model[39,40] (Fig.2b). The force applied by the tip is kept constant by a feedback loop, so the contact can also be effectively described as a flat punch determined only by the constant contact radius α (Fig.2c), measured experimentally. The tip-surface contact also acts as two springs in series, corresponding to the tip and the sample respectively. The effective Young´s modulus of the tip-surface system, $E^*$, is therefore given by

$$\frac{1}{E^*} = \frac{(1 - v_s^2)}{E_s} + \frac{(1 - v_{tip}^2)}{E_{tip}} \quad (1)$$

where $E_s$, $E_{tip}$, $v_s$ and $v_{tip}$ are the Young's modulus and the Poisson ratios of the sample and the tip respectively. $E^*$ is related to the contact stiffness $k^*$ as

$$E^* = \frac{k^*}{2a} \quad (2)$$

$k^*$ is the elastic constant of the spring that represents the tip-sample interaction[41], also known as contact stiffness and, like $E^*$, $k^*$ depends on the stiffness of the tip ($k_{tip}$), the stiffness of the sample ($k_s$).

The quantity we are after is the Young's modulus of the sample ($E_s$), which we could in principle calculate substracting $E_{tip}$ from $E^*$ in Eq.1. The problem is that we know neither $E^*$ nor $E_{tip}$, so we have one equation (Eq.1) with three unknowns. To solve this problem, we (i) measure the resonance frequencies of the cantilever and relate them to $E^*$ via elastic theory and (ii) measure the mechanical response of a part of the sample for which $E_s$ is known (in our case, the c-oriented $BaTiO_3$ domains), and use it for callibration. Knowing $E^*$ from and $E_s$ allows us to extract $E_{tip}$ and then repeat the analysis on the part of sample for which $E_s$ is unknown –the domain walls.

Based on the models of Hurley[42] and Rabe et al[43] the cantilever is modelled as a beam with length L, width w and thickness b, density ρ and Young's modulus $E_{can}$. The tip is located at a distance $L_1<L$ from the clamped end of the cantilever and the remaining distance to the other end of the cantilever is L´ (Fig 2a)[42]. The spring constant of the cantilever is calibrated experimentally by measuring force-displacement curves and the free resonance frequency (first harmonic, $\mathbf{f_1^0}$) of the cantilever. These parameters are shown in Table I.

Table I. Geometrical characteristics of the cantilever, experimental values of free resonance frequency and contact resonance frequencies for a and c domains of bulk $BaTiO_3$ and the corresponding wavelengths as calculated by equations 1, 2. Experimental value of cantilever's spring constant ($k_{lever}$) and the calculated normalized contact stiffness $k^*/k_{lever}$ of the system

| L(μm) | b(μm) | ρ(g/cm³) | $L_1$(μm) | $f_1^0$(kHz) | $E_{can}$(GPa) | $k_{lever}$(N/m) | α(nm) |
|---|---|---|---|---|---|---|---|
| 220 | 6,5 | 2,33 | 211,2 | 158 | 170 | 38 | 7 |

The experimental values of free resonance frequency ($f_1^{\ 0}$) (i.e. the resonance frequency when the tip of the cantilever is suspended above the sample without touching it) and the contact resonance frequency ($f_1$) (i.e. the resonance frequency when the tip is in contact with the sample) are used to relate the corresponding wavenumbers ($x_i$) through the equation[42]

$$x_1 L = x_1^0 L \sqrt{\frac{f_1}{f_1^0}} \quad (3)$$

Where the free cantilever wave number $x_1^0$ is given by[42]

$$(x_1^0 L)^2 = 4\pi f_1^0 \frac{L^2}{b}\sqrt{\frac{3\rho}{E_{can}}} \tag{4}$$

The normalized contact stiffness $k^*/k_{lever}$ is then given by[44,30]

$$\frac{k^*}{k_{lever}} = \frac{2}{3}(x_1 L_1)^3 \frac{(1 + cosx_1 L coshx_1 L)}{D} \tag{5}$$

$$D = (sinx_1 L' coshx_1 L' - cosx_1 L' sinhx_1 L')(1 - cosx_1 L_1 coshx_1 L_1) \\ - (sinx_1 L_1 coshx_1 L_1 - cosx_1 L_1 sinhx_1 L_1) \\ (1 + cosx_1 L' coshx_1 L') \tag{6}$$

The equations above connect the contact resonance frequencies and cantilever's spring constant to the contact stiffness $k^*$ and hence, using Eq.2, to the contact's effective Young's modulus $E^*$. If the sample's Young's modulus $E_s$ is also known (as it is for c-oriented barium titanate [45]), we can now use Eq. 1 to calculate the tip's Young's modulus ($E_{tip}$). The calculation can then be repeated on the domain wall, where $E_s$ is not known but $E_{tip}$ is. The results of the calculations are shown in Table II.

Table II. Young's modulus of BaTiO$_3$ based on literature[45] ($E_s$) and experimental results for c domains of BaTiO$_3$ as used to calibrate the Young's modulus of the tip ($E_{tip}$). For domain walls, all the experimental values and the derived values of Young's modulus.

|  | $f_1$(kHz) | $k^*/k_{lever}$ | $v_s$ | $E^*$(GPa) | $E_s$(GPa) |
|---|---|---|---|---|---|
| *c domains* | 738,5 | 239,92 | 0,3 | 17,07 | 63,6 |
| *domain walls* | 737 | 239,48 | 0,25 | 0,45 | 51,2 |

Considering the experimental results and using the model described above, the shift in resonance frequency of domain walls corresponds to a reduction of the effective Young's modulus of ~19%, with respect to the Young's modulus of the c domains.

## IV. Theory of ferroelectric domain wall softening

Having determined that the domain walls are mechanically softer than the domains despite being ferroelectric and not ferroelastic, the next question is why. In their seminal work, Tsuji et al[24] put forward three hypotheses: (i) defects, which are known to be attracted to domain walls, (ii) dynamic softening due to ferroelectric switching near the wall, and (iii) reduced depolarization energy at the domain wall, where there is no piezoelectricity. Let us examine these possibilities.

Defects are sample-dependent and common ones, such as oxygen vacancies, are notoriously difficult to quantify. The weakening of inter-atomic bonds caused by a vacancy should be fairly isotropic or at least orthotropic in the nearly cubic perovskite structure. That is to say, the defect-induced softening of the 180º walls should be similar in the in-plane and out-of-plane directions. We have compared the mechanical contrast of 180º walls inside the a-domains (polarization in-plane) with those in the c-domains (polarization out-of-plane) for the BaTiO$_3$ crystal, where both polarizations are accessible in a single scan due to the existence of a-c twins as well as 180º walls (figure 3-a). As Figure 3-b shows, while for c-domains (bubble domains) the 180º walls are softer, when the bubble domains penetrate into the a-domains (in-plane polarization) the mechanical contrast of the 180º walls disappears. The disappearance of mechanical contrast when the polarization is in-plane, combined with the fact that we observe the softening of out-of-plane walls in materials with different chemistries, leads us to believe that the role of chemical defects is less important than the out-of-plane orientation of the polarization.

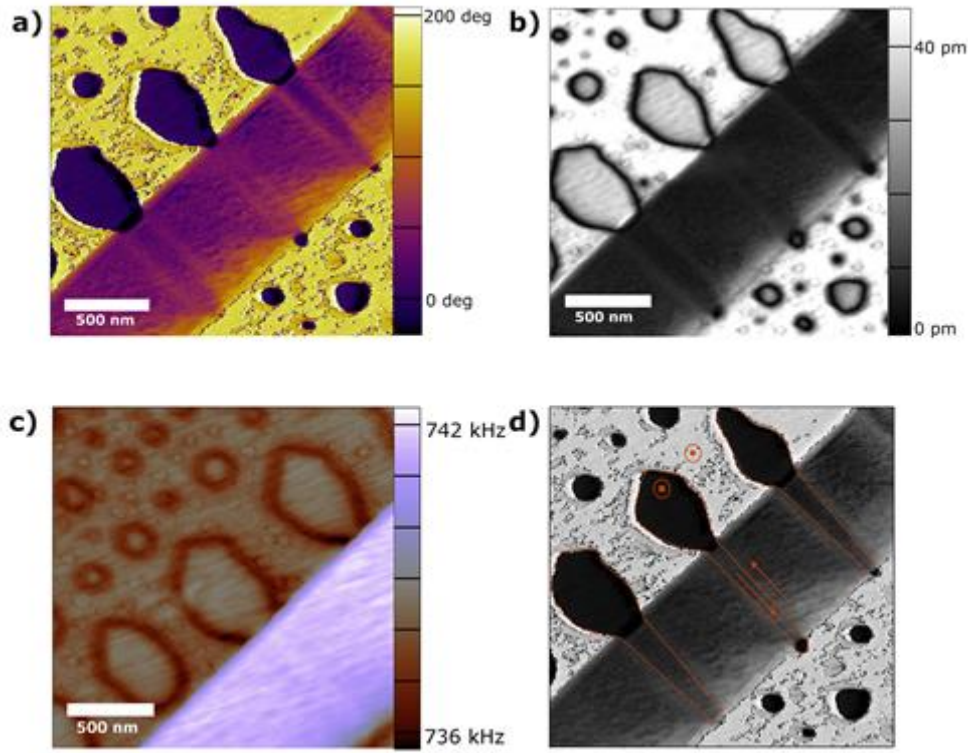

**Figure 3**. (a) Vertical PFM phase of BaTiO$_3$ single crystal spontaneously polarized, where the opposite out of plane polarization of the crystal is shown. (b) Vertical PFM amplitude of the crystal, where the in plane polarization is denoted(c)CR-FM image of the same area where there is no difference in frequency contrast due to in plane polarization. (d) Schematic representation of in plane polarization

The strain fields of a multidomain ferroelectric under the tip pressure is a mesoscopic problem too challenging for first-principles atomistic calculations. Instead we resort to a continuum model, with parameters for BaTiO$_3$ determined by previous first-principles work[46,47]. The starting point is the free energy density of the system, which can be described by the Ginzburg-Landau-Devonshire model[46,47],

$$f = f_l + f_g + f_q + f_c + f_f + f_{electr} \qquad (7)$$

$$f_l = a_{ij}P_iP_j + a_{ijkl}P_iP_jP_kP_l + a_{ijklmn}P_iP_jP_kP_lP_mP_n \tag{8}$$

$$f_g = G_{ijkl}\nabla_i P_j \nabla_k P_l/2 \tag{9}$$

$$f_q = -q_{ijkl}P_iP_j\varepsilon_{kl} \tag{10}$$

$$f_c = C_{ijkl}\varepsilon_{ij}\varepsilon_{kl}/2 \tag{11}$$

$$f_f = \frac{\Gamma_{ijkl}}{2}(\nabla_i P_j \varepsilon_{kl} - P_i \nabla_j \varepsilon_{kl}) \tag{12}$$

$$f_{electr} = \epsilon_r E^2/2 \tag{13}$$

where $f_l$ is the Landau free energy density for uniform ferroelectric polarization $P$, $f_g$ describes energy penalty for spatial variations of $P$, $f_q$ describes the interaction between the polarization and strains $\varepsilon_{ij}$ (electrostriction), $f_c$ is the elastic free-energy density, while $f_f$ denotes the contribution from flexoelectricity, the interaction between strain gradients and the polarization. $\Gamma_{ijkl}$ is the flexoelectric tensor. The strain $\varepsilon_{ij}$ is defined as $\frac{1}{2}(\nabla_i u_j + \nabla_j u_i)$ where $u$ are the displacements; summation over repeated indices is implied.

The integral of the free energy density over the entire crystal is minimized in the equilibrium situation. The electrostriction term, $f_q$, generates a spontaneous tensile strain along the polar direction inside the ferroelectric domains. This tensile strain is locally reduced at the wall due to the absence of polarization, which leads to a depression in the surface centered at the wall, as demonstrated in Fig. 4(b). The compressive pressure from CR-FM tip will interact with this pre-existing compressive strain profile. The result will be that the wall will move towards the tip, so that the domain wall depression coincides with the locum of the tip compression.

Pinning of the domain wall by the disorder potential and Peierls-Nabarro barriers results in a complex response. In the strong pinning regime, the wall will only bend towards the tip. This effect can be qualitatively captured by a simple free energy expansion, with a flat DW interacting with a parabolic pinning potential (second term), and the tip located at $x_{tip}$ and applying the force $F_z$,

$$E = E_0 - F_z u_z(x_{tip} - x_{DW}) + \frac{m\omega^2 x_{DW}^2}{2} \tag{14}$$

Expanding the surface profile in small $x_{DW}$, and minimizing the energy with regards to $x_{DW}$, we obtain $x_{DW} = -Fu'(x_{tip})/m\omega^2$, to the first order in $F$. The correction to the compliance is then $\Delta c = u'(x_{tip})^2/(m\omega^2)^2$. Hence, for significant softening, it is crucial that $u'(x_{tip})$ is large, leading to an increased effect when pressing within the DW strain footprint.

If the force applied by the tip is large enough to overcome the pinning potential, the wall will slide towards the tip, leading to a strongly nonlinear effect. These domain wall sliding modes usually have frequency in the GHz range[48–50]. As the domain wall's "strain-hollow" slides towards the AFM tip, the AFM will register a relatively large deformation in response to the stress, and thus a low effective stiffness (see Fig.4). To quantify this effect, it is necessary to solve the free energy in Eqs (7-13), which is analytically intractable but can be numerically computed by finite elements. We performed finite element simulations using known parameters from previous first-principles studies of $BaTiO_3$[46]. The contact of the CR-FM tip with the surface is simulated by applying a bell-shaped force $\propto a \exp\left(-\frac{x^2}{d^2}\right)$, with $2d \sim 20$ nm representing the diameter of the contact area of the tip, and $a=10^{-7}$ J/m$^3$. The softening was estimated as the ratio of maximum deformation induced by the tip at the wall and in the domain. In order to stay within a linear regime and avoid polarization domain switching, the virtual force applied to the tip was kept very low, less than a femto-Newton.

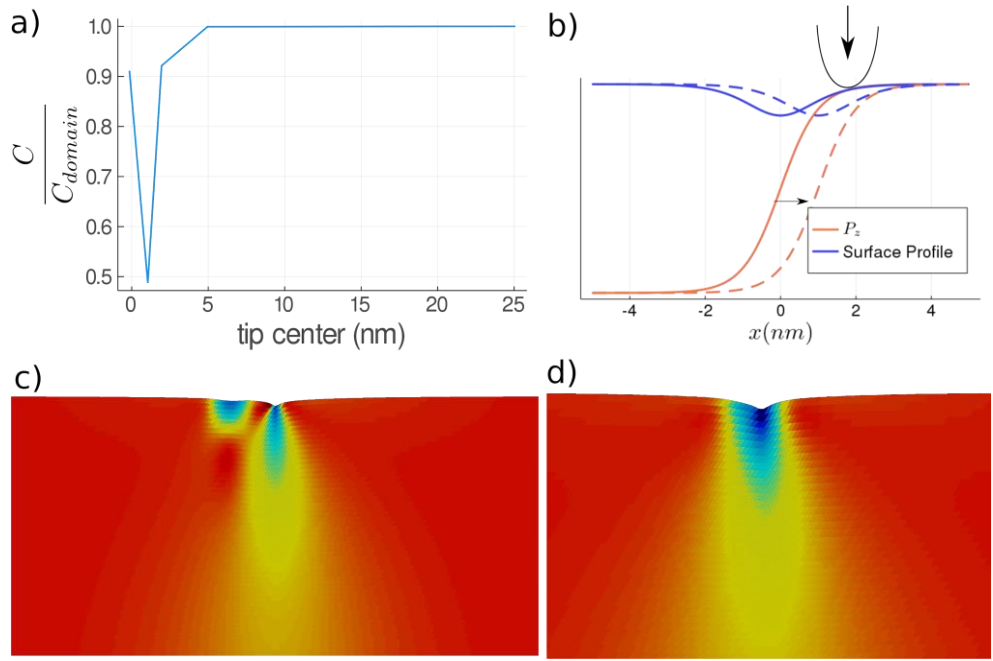

**Figure 4**. (a) Simulated change of stiffness as a function of distance between the CR-FM tip and a domain wall. The softening is maximized slightly away from the domain wall, but within its strain footprint, where the DW sliding mode contribution is important. Experimental situation corresponds to averaging within the tip region. (b) Schematic representation of the sliding mode. The polarization and strain profiles are shifted by dx. (c,d) Simulated strain profiles shown on a slice of the sample when the tip is near the DW (c) or further away (d). The dashed vertical line shows the initial position of the ferroelectric DW. The wall slides towards the tip in (c), as shown by the curved dashed line, whereas it does not move in (d).

These calculations predict that the elastic cost of deformation is smallest not at the domain wall itself, which is already spontaneously compressed and thus it is hard to compress further, but adjacent to it, where compression is achieved by the sliding of the domain wall (and its accompanying depression) towards the tip (Fig. 4a). As the tip moves further away, the distance becomes big enough that the stress field of the tip does not interact with the wall and the material recovers its intrinsic stiffness.

Another potential source of mechanical contrast is depolarization. Applying tip pressure to the surface of a piezoelectric (all ferroelectrics are piezoelectric) by definition modifies its polarization and thus has an electrostatic energy cost. The AFM tip induces deformations $\varepsilon_{ij}$ that are inhomogeneous (large near the tip, small far from the tip), so the polarization due piezoelectricity, $P_i \sim e_{ijk}\varepsilon_{jk}$, is not homogeneous. Tip pressure therefore induces bound charges $\nabla \cdot P \neq 0$ in the inhomogeneously deformed region, and these create a depolarizing field. Higher depolarization implies bigger work and thus higher effective stiffness.

We identify two main mechanisms of formation of bound charges under the tip: 1) Variation of in-plane polarization, $\nabla_1 P_1 \neq 0$, induced by shear piezoelectricity $P_1 \sim e_{15}\varepsilon_5$ (Fig. 5a); 2) Generation of out-of-plane polarization due to longitudinal piezoelectricity, $\Delta P_3 \sim e_{11}\varepsilon_3$, which is unscreened in the case of open boundary conditions and screened at short-circuit conditions (Fig. 5b).

Because all the piezoelectric constants $e_{jm}$ flip their signs across the domain wall, $e_{jm}(P_3 \downarrow) = -e_{jm}(P_3 \uparrow)$, the distribution of the tip-induced bound charges is qualitatively different when the tip is pressed at the domain and at the domain wall. For the charges induced by the "in-plane" mechanism (Fig. 5a), the in-plane polarization $P_1 \sim e_{15}\varepsilon_5$ forms head-to-head or tail-to-tail pattern with corresponding bound charges when the tip is pressed in the domain, and head-to-tail pattern when the tip is pressed at the domain wall, implying significantly reduced electrostatic energy costs and softer mechanical response in the latter case. Likewise, the surface charges generated by the "out-of-plane" mechanism (Fig. 5b) have monopole-like distribution when the tip is pressed in the domain and dipole-like distribution when the tip is pressed at the domain wall, again implying lower electrostatic energy costs and a softer mechanical response of the domain wall at open electric boundary conditions. Notice that this polarity-dependent orientation of the piezoelectric response is qualitatively different from flexoelectricity, which is polarity-independent and thus less sensitive to the presence of a polar domain wall; for this reason, we have discarded flexoelectricity from the analysis.

We have tested the reasoning given above by modelling a simplified two-dimensional system, with the contact between the CR-FM tip and the surface described by an out-of-plane force proportional to $e^{-\frac{x^2}{t^2}}$, where $2t \sim 20$ nm is the contact area. As in the experiment, we focus only on linear static elastic effects and apply a force small enough to avoid any polarization switching. All the simulations are done with two electric boundary conditions: open boundary conditions with surface screening of the polarization by immobile surface charges (requiring the normal component of the electric displacement field at the surface $D_n = \epsilon_r E_n + P_n = P_S \tanh x/\xi$ at all times, where $\xi \sim 1$ nm), and short-circuit boundary conditions. We note that the experimentally investigated crystal is closer to the short-circuit case in the out-of-plane direction: even though the film surface is not electroded, the AFM tip in contact with the surface is metallic.

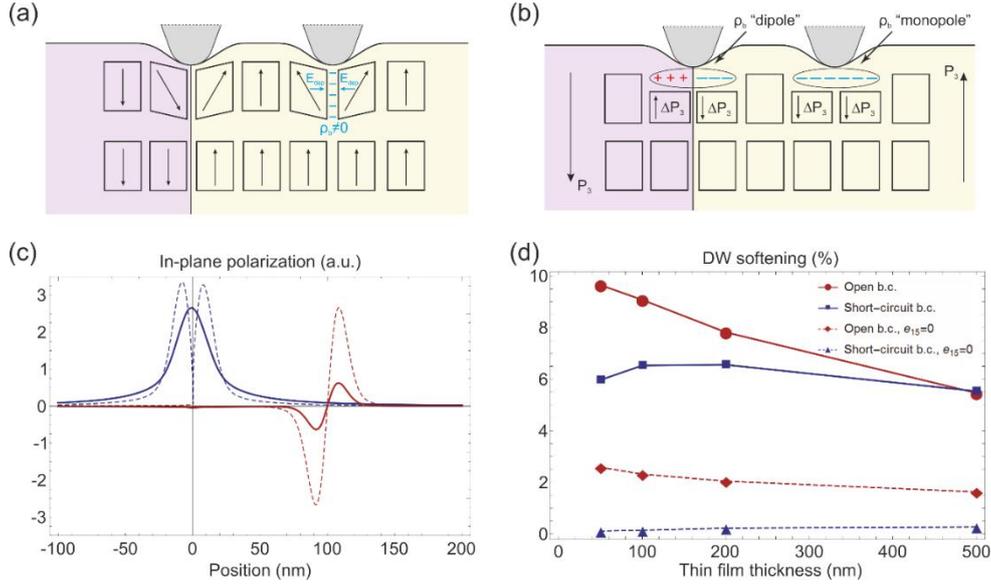

**Figure 5**. (a) Schematic of the "in-plane" mechanism of the bound charge formation: applying the tip to the surface induces in-plane polarization via the shear piezoelectricity: $P_1 \sim e_{15}\varepsilon_5$, with larger depolarizing electric fields $E_{dep}$ when tip is applied in the bulk domain. (b) Schematic of the "out-of-plane" mechanism of the domain wall softening: applying the tip to the surface induces out-of-plane polarization $\Delta P_3 \sim e_{13}\varepsilon_1 + e_{33}\varepsilon_3$. (c) Manifestation of the depolarizing effect on the tip-induced in-plane polarization $P_1$. Solid lines: polarization extracted from finite-element modeling at distance 5nm from the film surface, with the tip applied at the domain wall (blue) and in the bulk domain (red). Dashed lines: polarization expected due to the shear piezoelectric effect $P_1 \approx e_{15}\varepsilon_5$, with shear strain $\varepsilon_5$ extracted at distance 5nm from the film surface. (d) Study of the domain wall softening at different electric boundary conditions, values of the shear piezoelectric coefficient $e_{15}$ and film thicknesses.

In all studied cases, we obtain that, when the AFM tip is applied at the domain wall, the in-plane polarization $P_1$ generated under the tip (blue solid line in Fig. 5c) is described with good precision by the piezoelectric effect $P_1 \approx e_{15}\varepsilon_5$ (blue dashed line in Fig. 5c). On the other hand, when the tip is applied in the bulk domain, the generated in-plane polarization is strikingly smaller than the one expected from piezoelectricity (Fig. 5c, red lines). This suppression of the in-plane polarization is due to the depolarizing cost of the head-to-head configuration and, as expected, is accompanied by a harder elastic response of the bulk domain compared to the domain wall. This translates into an apparent DW softening of 5-10% (Fig. 5d) relative to the stiffness of the domain.

To confirm the link between the suppression of in-plane polarization and elastic hardening of the bulk, we have also performed an additional simulation with the shear piezoelectric constant $e_{15} = 0$ (by setting the shear electrostriction coefficient $q_{44} = 0$ in Eq. (10)) at *short-circuit* boundary conditions, thus removing both the bound charges appearing "in-plane" and "out-of-plane". As a result, the DW softening almost vanishes, down to <0.3% (Fig. 5d). Then, reactivating the "in-plane" mechanism of charge formation by returning $e_{15}$ to the BaTiO$_3$ value leads to an averaged 6% increase of the DW softening, and applying open boundary conditions increases the DW softening by another 2% (Fig. 5d). The somewhat surprising conclusion, therefore, is that the main contributor to the electrostatic softening at the wall is NOT the out-of-plane piezoelectricity but the in-plane (shear) piezoelectricity. This is important because, while the former can be partially screened by the use of metallic tips or in the presence of electrodes or adsorbates, the latter cannot. The in-plane piezoelectric contribution to domain wall softening is therefore unavoidable, despite the polar axis being out-of-plane.

V.     Conclusions

Although ferroelectric 180º domains are mechanically identical, the domain walls that separate them display a mechanical contrast: the walls are softer than the domains. This reduced effective stiffness is detected as a lowering of the mechanical resonance frequency of the AFM cantilever in contact with the domain wall, an effect first observed in ceramics of lead zirconate-titanate[24,25] and reiterated here on single crystals of uniaxial lithium niobate and perovskite $BaTiO_3$, and epitaxial thin films of $PbTiO_3$. The effect therefore appears to be general.
Theoretical modeling shows that there can be at least two contributing factors to the vertical ease of deformation of the wall: domain wall sliding and depolarization-activated electromechanical coupling. For $BaTiO_3$, these two contributions have a quantitatively comparable impact on the total softening.

Ferroelectric 180º domain walls have, by definition, P=0 in the vertical direction, so the spontaneous strain associated with P is suppressed at the wall. When an inhomogeneous vertical compression is delivered by the AFM tip near the wall, therefore, the material can respond by sliding (or broadening) the wall so that this region of inherently reduced vertical strain (aka strain dip) moves under the tip. In this way, maximum deformation is achieved with minimum effort. This dynamic response requires the strain field from the tip to be inhomogeneous and asymmetrically located with respect to the wall; if there was the same amount of compression on either side of the wall, it would not move. However, it is important to emphasize that this effect is not flexoelectric.

The mechanical response is also related to the depolarization field generated upon straining a piezoelectric material (all ferroelectrics are piezoelectric). Tip-induced inhomogeneous deformation causes shear strain, thereby generating an in-plane piezoelectric polarization via shear piezoelectric effect. By symmetry, in a tetragonal ferroelectric this in-plane component must be head-to-head or tail-to-tail in the domains, whereas it is head-to-tail in the domain walls – hence, the electrostatic costs are smaller at the domain walls, which facilitates their deformation.

The fact that purely ferroelectric (i.e. non-ferroelastic) domain walls display mechanical contrast with respect to their surroundings has consequences not only for the mechanical properties of the material, but also, in principle, for any functional property linked to the material's elasticity. This includes heat transport, which is mediated by phonons and thus by lattice vibrations. As argued at the introduction, ferroelectric domain walls are known to affect heat transport[26] and have been proposed as the basis for phononic switches[27,28], and the mechanical contrast at the wall means that periodically-poled ferroelectric crystals can also be regarded as phononic crystals. A more general message is that domain walls are not only functionally distinct, but also mechanically singular, and a complete picture of domain wall physics must incorporate this mechanical singularity.

**ACKNOWLEDGEMENTS**

C.S. thanks BIST for the PREBIST Grant. This project has received funding from the European Union's Horizon 2020 research and innovation program under the Marie Skłodowska-Curie grant agreement No. 754558.

E.L. acknowledges the funding received from the European Union's Horizon 2020 research and innovation program through the Marie Skłodowska-Curie Actions: Individual Fellowship-Global Fellowship (ref. MSCA-IF-GF-708129).

M.S. and K.S. acknowledge the support of the European Research Council under the European Union's Horizon 2020 research and innovation program (Grant Agreement No. 724529), Ministerio de Economia, Industria y Competitividad through Grant Nos. MAT2016-77100-C2-2-P and SEV-2015-0496, and the Generalitat de Catalunya (Grant No. 2017SGR 1506).

**References.**

1. Seidel, J. *et al.* Conduction at domain walls in oxide multiferroics. *Nat. Mater.* **8**, 229–234 (2009).
2. Farokhipoor, S. & Noheda, B. Conduction through 71° Domain Walls in BiFeO3 Thin Films. *Phys. Rev. Lett.* **107**, 127601 (2011).
3. He, Q. *et al.* Magnetotransport at Domain Walls in BiFeO3. *Phys. Rev. Lett.* **108**, 67203 (2012).
4. Lee, J. H. *et al.* Spintronic Functionality of BiFeO3 Domain Walls. *Adv. Mater.* **26**, 7078–7082 (2014).
5. Murari, N. M., Hong, S., Lee, H. N. & Katiyar, R. S. Direct observation of fatigue in epitaxially grown Pb(Zr,Ti)O3 thin films using second harmonic piezoresponse force microscopy. *Appl. Phys. Lett.* **99**, 52904 (2011).
6. Aird, A. & Salje, E. K. H. Sheet superconductivity in twin walls: experimental evidence of WO3-x. *J. Phys. Condens. Matter* **10**, (1998).
7. Guyonnet, J., Gaponenko, I., Gariglio, S. & Paruch, P. Conduction at Domain Walls in Insulating Pb(Zr0.2Ti0.8)O3 Thin Films. *Adv. Mater.* **23**, 5377–5382 (2011).
8. Schröder, M. *et al.* Conducting Domain Walls in Lithium Niobate Single Crystals. *Adv. Funct. Mater.* **22**, 3936–3944 (2012).
9. Choi, T. *et al.* Insulating interlocked ferroelectric and structural antiphase domain walls in multiferroic YMnO3. *Nat Mater* **9**, 253–258 (2010).
10. McQuaid, R. G. P., Campbell, M. P., Whatmore, R. W., Kumar, A. & Marty Gregg, J. Injection and controlled motion of conducting domain walls in improper ferroelectric Cu-Cl boracite. *Nat. Commun.* **8**, 1–7 (2017).
11. Meier, D. Functional domain walls in multiferroics. *J. Phys. Condens. Matter* **27**, 463003 (2015).
12. Catalan, G., Seidel, J., Ramesh, R. & Scott, J. F. Domain wall nanoelectronics. *Rev. Mod. Phys.* **84**, 119–156 (2012).
13. Salje, E. K. H. Multiferroic Domain Boundaries as Active Memory Devices: Trajectories Towards Domain Boundary Engineering. *ChemPhysChem* **11**, 940–950 (2010).
14. Muñoz-Saldaña, J., Schneider, G. A. & Eng, L. M. Stress induced movement of ferroelastic domain walls in BaTiO3 single crystals evaluated by scanning force microscopy. *Surf. Sci.* **480**, (2001).
15. Anbusathaiah, V. *et al.* Labile ferroelastic nanodomains in bilayered ferroelectric thin films. *Adv. Mater.* **21**, 3497–3502 (2009).
16. Zhang, T. Y. & Gao, C. F. Fracture behaviors of piezoelectric materials. *Theor. Appl. Fract. Mech.* **41**, 339–379 (2004).
17. Abdollahi, A. & Arias, I. Phase-field modeling of crack propagation in piezoelectric and ferroelectric materials with different electromechanical crack conditions. *J. Mech. Phys. Solids* **60**, 2100–2126 (2012).
18. Park, S. M. *et al.* Selective control of multiple ferroelectric switching pathways using a trailing flexoelectric field. *Nat. Nanotechnol.* **13**, 366–370 (2018).
19. Shapovalov, K. *et al.* Elastic coupling between nonferroelastic domain walls. *Phys. Rev. Lett.* **113**, 1–5 (2014).
20. Johnson-Wilke, R. L. *et al.* Ferroelectric/Ferroelastic domain wall motion in dense and porous tetragonal lead zirconate titanate films. *IEEE Trans. Ultrason. Ferroelectr. Freq. Control* **62**, 46–55 (2015).

21. Anbusathaiah, V. *et al.* Ferroelastic domain wall dynamics in ferroelectric bilayers. *Acta Mater.* **58**, 5316–5325 (2010).
22. Xu, F. *et al.* Domain wall motion and its contribution to the dielectric and piezoelectric properties of lead zirconate titanate films. *J. Appl. Phys.* **89**, 1336–1348 (2001).
23. Damjanovic, D. & Demartin, M. Contribution of the irreversible displacement of domain walls to the piezoelectric effect in barium titanate and lead zirconate titanate ceramics. *J. Phys. Condens. Matter* **9**, 4943–4953 (1997).
24. Tsuji, T. *et al.* Significant stiffness reduction at ferroelectric domain boundary evaluated by ultrasonic atomic force microscopy. *Appl. Phys. Lett.* **87**, 71909 (2005).
25. Tsuji, T. *et al.* Evaluation of domain boundary of piezo/ferroelectric material by ultrasonic atomic force microscopy. *Japanese J. Appl. Physics, Part 1 Regul. Pap. Short Notes Rev. Pap.* **43**, 2907–2913 (2004).
26. Langenberg, E. *et al.* Ferroelectric Domain Walls in PbTiO3 Are Effective Regulators of Heat Flow at Room Temperature. *Nano Lett.* (2019) doi:10.1021/acs.nanolett.9b02991.
27. Seijas-Bellido, J. A. *et al.* A phononic switch based on ferroelectric domain walls. *Phys. Rev. B* **96**, 140101 (2017).
28. Ihlefeld, J. F. *et al.* Room-temperature voltage tunable phonon thermal conductivity via reconfigurable interfaces in ferroelectric thin films. *Nano Lett.* **15**, 1791–1795 (2015).
29. Ferraro, P., Grilli, S. & De Natale, P. Ferroelectric crystals for photonic applications: Including nanoscale fabrication and characterization techniques, second edition. *Springer Series in Materials Science* vol. 91 (2014).
30. Rabe, U. *et al.* Quantitative determination of contact stiffness using atomic force acoustic microscopy. *Ultrasonics* (2000) doi:10.1016/S0041-624X(99)00207-3.
31. Harnagea, C., Pignolet, A., Alexe, M. & Hesse, D. Piezoresponse Scanning Force Microscopy: What Quantitative Information Can We Really Get Out of Piezoresponse Measurements on Ferroelectric Thin Films. *Integr. Ferroelectr.* **44**, 113–124 (2002).
32. Soergel, E. Piezoresponse force microscopy (PFM). *J. Phys. D. Appl. Phys.* **44**, 464003 (2011).
33. Wei, X. K. *et al.* Néel-like domain walls in ferroelectric Pb(Zr,Ti)O 3 single crystals. *Nat. Commun.* **7**, 1–7 (2016).
34. De Luca, G. *et al.* Domain Wall Architecture in Tetragonal Ferroelectric Thin Films. *Adv. Mater.* **29**, 1–5 (2017).
35. Kim, S., Gopalan, V. & Gruverman, A. Coercive fields in ferroelectrics: A case study in lithium niobate and lithium tantalate. *Appl. Phys. Lett.* **80**, 2740–2742 (2002).
36. Cordero-Edwards, K., Domingo, N., Abdollahi, A., Sort, J. & Catalan, G. Ferroelectrics as Smart Mechanical Materials. *Adv. Mater.* 1702210-n/a (2017) doi:10.1002/adma.201702210.
37. Cordero-Edwards, K., Kianirad, H., Canalias, C., Sort, J. & Catalan, G. Flexoelectric Fracture-Ratchet Effect in Ferroelectrics. *Phys. Rev. Lett.* **122**, 135502 (2019).
38. Abdollahi, A. *et al.* Fracture toughening and toughness asymmetry induced by flexoelectricity. *Phys. Rev. B - Condens. Matter Mater. Phys.* **92**, 094101 (2015).
39. Rabe, U. *et al.* Imaging and measurement of local mechanical material properties by atomic force acoustic microscopy. *Surf. Interface Anal.* **33**, 65–70 (2002).

40. Oliver, W. C. & Brotzen, F. R. On the generality of the relationship among contact stiffness, contact area, and elastic modulus during indentation. *J. Mater. Res.* **7**, 613–617 (1992).
41. Rabe, U. & Arnold, W. Atomic force microscopy at MHz frequencies. *Ann. Phys.* **506**, 589–598 (1994).
42. Hurley, D. C. Contact Resonance Force Microscopy Techniques for Nanomechanical Measurements. in *Applied Scanning Probe Methods XI: Scanning Probe Microscopy Techniques* 97–138 (Springer Berlin Heidelberg, 2009). doi:10.1007/978-3-540-85037-3_5.
43. Rabe, U., Kopycinska-Müller, M. & Hirsekorn, S. Atomic Force Acoustic Microscopy. in *Acoustic Scanning Probe Microscopy* (eds. Marinello, F., Passeri, D. & Savio, E.) 123–153 (Springer Berlin Heidelberg, 2013). doi:10.1007/978-3-642-27494-7_5.
44. Turner, J. A., Hirsekorn, S., Rabe, U. & Arnold, W. High-frequency response of atomic-force microscope cantilevers. *J. Appl. Phys.* **82**, 966–979 (1997).
45. Berlincourt, D. & Jaffe, H. Elastic and Piezoelectric Coefficients of Single-Crystal Barium Titanate. *Phys. Rev.* **111**, 143–148 (1958).
46. Marton, P., Rychetsky, I. & Hlinka, J. Domain walls of ferroelectric BaTiO3 within the Ginzburg-Landau-Devonshire phenomenological model. (2010) doi:10.1103/PhysRevB.81.144125.
47. Hlinka, J. & Márton, P. Phenomenological model of a 90° domain wall in BaTiO3-type ferroelectrics. *Phys. Rev. B - Condens. Matter Mater. Phys.* **74**, 1–12 (2006).
48. Zheng, L. *et al.* Low-energy structural dynamics of ferroelectric domain walls in hexagonal rare-earth manganites. *Sci. Adv.* **3**, e1602371 (2017).
49. Hlinka, J., Paściak, M., Körbel, S. & Marton, P. Terahertz-Range Polar Modes in Domain-Engineered BiFeO3. *Phys. Rev. Lett.* **119**, 1–6 (2017).
50. Prosandeev, S., Yang, Y., Paillard, C. & Bellaiche, L. Displacement Current in Domain Walls of Bismuth Ferrite. *npj Comput. Mater.* **4**, 1–9 (2018).